# Solvent constraints for biopolymer folding and evolution in extraterrestrial environments

Ignacio E. Sánchez*, Ezequiel A. Galpern and Diego U. Ferreiro*

Laboratorio de Fisiología de Proteínas, Facultad de Ciencias Exactas y Naturales, Universidad de Buenos Aires and Consejo Nacional de Investigaciones Científicas y Técnicas, Instituto de Química Biológica de la Facultad de Ciencias Exactas y Naturales (IQUIBICEN-CONICET), Buenos Aires CP1428, Argentina.

* To whom correspondence may be addressed: isanchez@qb.fcen.uba.ar, ferreiro@qb.fcen.uba.ar

## Abstract

We propose that spontaneous folding and molecular evolution of biopolymers are two universal aspects that must concur for life to happen. These aspects are fundamentally related to the chemical composition of biopolymers and crucially depend on the solvent in which they are embedded. We show that molecular information theory and energy landscape theory allow us to explore the limits that solvents impose on biopolymer existence. We consider 54 solvents, including water, alcohols, hydrocarbons, halogenated solvents, aromatic solvents, and low molecular weight substances made up of elements abundant in the universe, which may potentially take part in alternative biochemistries. We find that along with water, there are many solvents for which the liquid regime is compatible with biopolymer folding and evolution. We present a ranking of the solvents in terms of biopolymer compatibility. Many of these solvents have been found in molecular clouds or may be expected to occur in extrasolar planets.

**Introduction**

Biology is based on biochemistry and biochemistry as we know it requires biopolymers. There are two fundamental characteristics that can be expected to be universal for any biopolymer: their potential for folding and evolution. It has been argued that biopolymers allow processes of folding and assembly to be detached from the required investment of free energy (1). For small molecules, by contrast, assembly and investment are directly coupled. Therefore, small molecules cannot achieve the elaborate folds and assemblies that we find on terrestrial biological relics. Spontaneous folding to specific structures is one of the fundamental aspects that come naturally to biopolymers. Specific, fast and robust folding can only occur in polymers that follow the 'Principle of Minimal Frustration' (2), for which there is a fundamental correlation between similarity in structure and energetic distribution. In these polymers, the folded structure can be encoded in the sequence of monomers given an environment (3). The application of statistical mechanics to polymer models define the parameters relevant to describe the characteristic structural transitions of these systems (4). Quantitative molecular interpretation of these theories provide the basis for the exploration of the ranges of the constraints that are expected to be found for any biological polymer.

Besides (but not independently of) folding, biopolymers must be able to change over time, they must explore new structural and functional forms that allow for biochemistry to evolve. Biopolymer evolution can be fundamentally described as explorations on the sequence space that encodes the structural forms (5). Changes in the sequences relate to changes in the structures, giving rise to genotype to phenotype mappings. For terrestrial biopolymers, this mapping is certainly complex and depends on the details of the environment in which the information is decoded. Recently, molecular information theory was conjured with energy landscape theory to find the evolutionary informational footprint in foldable polymers (6). It was found that the average information contained in the sequences of evolved terrestrial proteins is very close to the average information needed to specify a fold. Moreover, it was shown that it is possible to compute the efficiency for conversion of folding free energy into sequence information, that for terrestrial proteins is around 50%.

We propose here that spontaneous folding and molecular evolution are two universal aspects that must concur whether life is happening. These aspects are fundamentally related to the chemical composition of the biopolymers and crucially depend on the solvent in which these are embedded. It is generally accepted that life must occur in liquid solvents, as this is the phase in which relevant biochemistry can happen. Some characteristics that are expected for possible alternative solvents are the ability to form hydrogen bonds, the presence of hydrophobic phase separation allowing for membrane systems, acid/base properties, chemical stability of known biomolecules, feasibility of a wide range of metabolic reactions, temperature range of liquid, etc (1–3). There is a wide range of proposed life-sustaining solvents that includes water, ammonia, sulfuric acid, formamide, hydrocarbons, dihydrogen, dinitrogen, carbon dioxide, carbon disulfide, hydrazine, hydrogen cyanide, hydrogen sulfide, nitric oxide, silicon dioxide to name some (1–3). Here, we abstract out the detailed chemistry of the polymer and concentrate on the solvent characteristics that constrain biopolymer folding and evolution in extraterrestrial environments.

**Theory**

**Characteristic temperatures of biopolymer folding and evolution**

Energy landscape theory and molecular information theory deduce in an abstract manner the conditions that are necessary for the evolution of folded biopolymers. Since the deductions do not depend on the chemical nature of the biopolymer or the solvent, the results can be used to assess whether alternative biochemistries support biopolymer existence.

Energy landscape theory and molecular information theory identify at least four characteristic temperatures of biopolymer folding and evolution (6). First, the physiological temperature $T_{phys}$ is the one at which the molecule functions in a biological setting. Second, the folding temperature $T_f$ is the one at which 50% of the molecules remain folded. Empirically, on earth, $T_f$ is on average around 15 degrees higher than $T_{phys}$ (6). Given the broad temperature ranges considered in this work, we will approximate $T_{phys}=T_f$ for simplicity. Third, the glass temperature $T_g$ is related to the thermodynamics of trapping the polymer in the configurational space. Below $T_g$, the system runs out of entropy and its kinetics exhibit unfoldable glass-like behavior (2). Fourth, the selection temperature $T_{sel}$ is related to the strength of selection for folding stability during evolution of the biopolymer sequence in the sequence space. If $T_{sel}$ is lower than $T_g$, a biopolymer sequence can evolve for folding.

We propose that evolved biopolymer sequences must correspond with polymers that are folded and active at the physiological temperature. From energy landscape theory considerations, the relationship between these four characteristic temperatures is (6):

$$T_{sel}<T_g<T_{phys}<T_f \quad [1]$$

Equation [1] implies that the ratios $T_g/T_{sel}$ and $T_f/T_g$ are both higher than one. Both conditions need to be met for evolution of folded biopolymers. The condition for $T_g/T_{sel}$ ensures that foldable biopolymers' sequences can evolve in a relevant timescale, while the condition for $T_f/T_g$ ensures that evolved biopolymer structure can fold in a relevant timescale.

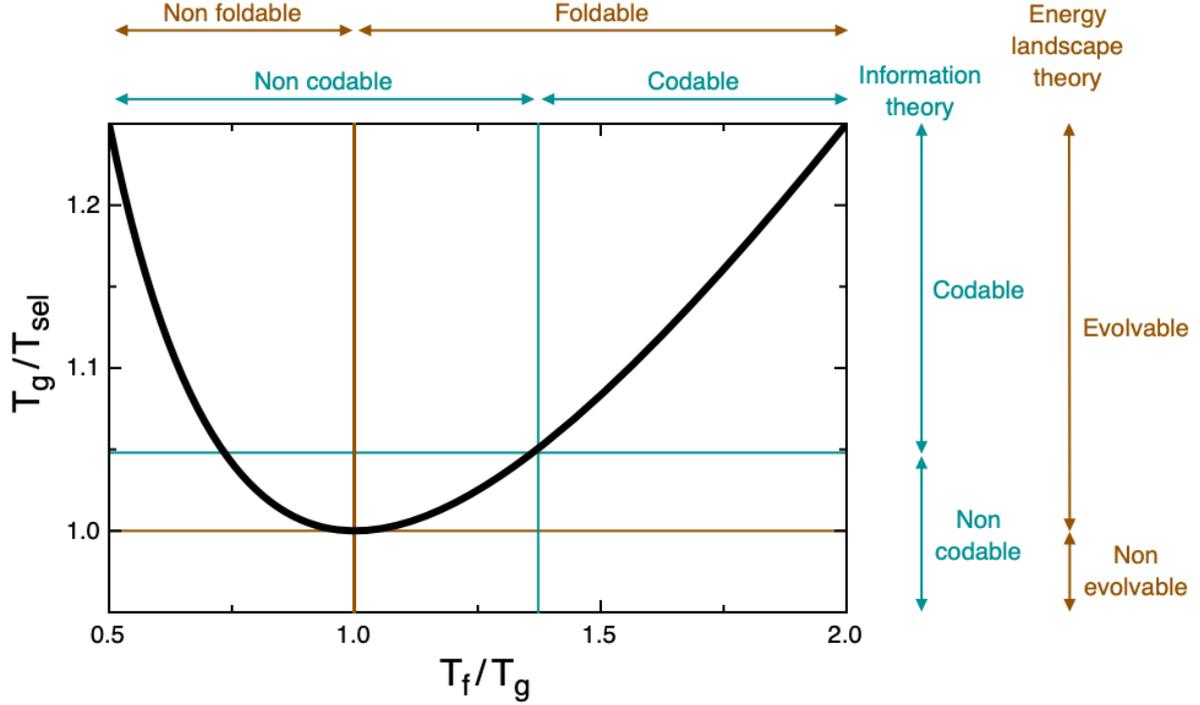

*Figure 1. Relationship between the characteristic temperatures for folding and evolution of biopolymers.* The $T_f/T_g$ ratio indicates the funneling of the folding energy landscape, which directly relates to the foldability of a biopolymer. The $T_g/T_{sel}$ ratio relates to the selection strength of the sequences that code for a biopolymer. The limits imposed by energy landscape theory are shown in brown and the limits set by molecular information theory in blue.

We can refine these theoretical limits given by equation [1] as follows. Energy landscape theory provides the following relationship between $T_{sel}$, $T_g$ and $T_f$ (6–8):

$$\frac{2}{T_f T_{sel}} = \frac{1}{T_g^2} + \frac{1}{T_f^2} \qquad [2]$$

According to molecular information theory, the folded structures can be encoded in sequences of monomers, and the ratio of $T_{sel}$ to $T_{phys}$ defines the efficiency for conversion of folding free energy into sequence information. This ratio has an upper limit of $ln(2)$ (6, 9):

$$Eff = \frac{T_{sel}}{T_{phys}} \simeq \frac{T_{sel}}{T_f} \leq ln(2) \qquad [3]$$

If we solve equation equation [2] for $T_g$ and substitute $T_{sel} = Eff \cdot T_f$, we obtain

$$T_f/T_g = \sqrt{(2 - Eff)/Eff} \geq 1.373 \qquad [4]$$

If we solve equation [2] for $T_f$ and combine the solution with equation [3] we obtain $Eff = (T_{sel}/T_g)^2/(1 + \sqrt{1 - (T_{sel}/T_g)^2})$. Since $Eff \leq ln(2)$, it follows that

$$T_g/T_{sel} \geq 1.048 \qquad [5]$$

Finally, putting together [4] and [5] we get

$$T_f/T_{sel} \geq 1.439 \qquad [6]$$

Since equation [2] comes from a simplified random energy model for the energetics of folding and evolution, the precise values obtained should be considered as first order approximations. The combination of energy landscape theory with molecular information theory provides new theoretical limits for the ratios between $T_{sel}$, $T_g$ and $T_f$. An evolved, coded, biopolymer able to fold into a specific conformation should fulfill conditions [4] to [6]. A general diagram of these relations is presented in Figure 1. The $T_f/T_g$ ratio indicates the funneling of the folding energy landscape, which directly relates to the foldability of the biopolymer (4). The $T_g/T_{sel}$ ratio is related to the selection strength of the sequences, as the ratio increases the possibility of finding codable sequences by mere chance decreases, as the sequence entropy of the foldable ensemble must decrease as $T_{sel}$ lowers. Notice that there is a fundamental compromise between foldability and evolvability: the better the folding, the more difficult it is to find sequences that will code for it (10). This result is independent of the chemistry of the polymer and the physics of the solvent, and we propose that this is a universal characteristic expected to occur in biological polymers in any extraterrestrial environment.

**Estimation of $T_g/T_s$, $T_f/T_g$ and efficiency**

Here we estimate $T_g/T_{sel}$, $T_f/T_g$ and efficiency for folding of biopolymers in non-aqueous solvents. We estimate $T_g$ from the glass transition temperatures of known terrestrial biopolymers (see below), and $T_f$ from the range of temperatures in which a solvent is in the liquid state. From this data, we can directly calculate the possible values for the $T_f/T_g$ ratio. We can also calculate the efficiency for conversion of free energy into information by evaluating equation [16] from (6) for the case $T_{phys}=T_f$:

$$Eff = 2/(1 + (T_f/T_g)^2) \qquad [7]$$

Finally, the $T_{sel}$ can be calculated by combining equations [3] and [7]. It follows that the ratio $T_g/T_{sel}$ is:

$$T_g/T_{sel} = (T_g/T_f)(1 + (T_f/T_g)^2)/2 \qquad [8]$$

The ratio $T_g/T_{sel}$ has a minimum at $T_f=T_g$ (Figure 1). As a consequence, the ratio $T_g/T_s$ can potentially be above the theoretical minimum in one or two temperature ranges. The lower temperature range for which $T_g/T_{sel}$ is above the theoretical minimum corresponds with values of $T_f/T_g$ below 1 and efficiencies above 1. This suggests a scenario in which biopolymer sequences can evolve but not fold in a relevant timescale. The higher temperature range for which $T_g/T_{sel}$ is above the theoretical minimum (foldable biopolymer sequences can be coded and evolve in a relevant timescale) corresponds with values of $T_f/T_g$ above the theoretical minimum and efficiencies below the theoretical maximum (evolved biopolymer sequences can fold in a relevant timescale). We consider this the biologically relevant range for the $T_g/T_{sel}$ ratio.

## Results

**Biopolymer folding in water**

In this section we examine the properties of water to provide values for $T_f$ and $T_g$. As a starting point, we focus on a pressure of 1 atm. The potential influence of changes in environmental pressure is discussed in a latter section. Under these conditions, water is a liquid from 273.15 to 373.13 K. Since terrestrial life mainly exists in liquid water, we assume that $T_f$ may take values in this range of temperatures.

Next we consider the value of the glass temperature $T_g$ for biopolymer folding in water. Water is a complex substance presenting multiple transitions (11) (Figure 2, left). Liquid water can be superheated to about 553 K, and small droplets with 1 to 10 microns in diameter can be supercooled to about 231 K (11). Water, like any other liquid, can be vitrified when cooled fast enough to avoid crystallization. Most experimental observations report that rapidly cooled water at atmospheric pressure has a glass transition temperature of about 136 K (11). The dynamic and transport properties of supercooled water show critical behavior as a function of temperature, as indicated by various techniques such as viscosity, dielectric relaxation and NMR measurements, neutron and light scattering spectroscopies and molecular dynamics simulations (11–13). The estimated values for the critical temperature of supercooled water range from 212 to 228 K, with an average of 220±7 K (11–13) (Figure 1, red horizontal line).

Biopolymer folding and solvent dynamics are intimately linked. Large-scale protein motions follow the solvent fluctuations, but can be slower by a large factor. Slowing takes place because large-scale motions consist of many small steps, each determined by the rate constant for solvent fluctuations. This phenomenon is called slaving (14) and suggests that a critical temperature for a solvent transition may be considered an upper limit for the glass transition of the biopolymers. As discussed above, this temperature may be 136 K or close to 220 K in the case of water, depending on which solvent transition is most relevant for folding. We can consider whether this is the case for three terrestrial biopolymers undergoing structural transitions in aqueous media: proteins, RNA and DNA, which were studied using molecular dynamics simulations and neutron scattering (Figure 2, right). The glass transition of model proteins takes place between 180 and 210 K, with an average value of 196±14 K (15–19) (Figure 2, blue circles). The glass transition of different RNA samples takes place between 190 and 230 K, with an average value of 212±17 K (16, 20–22) (Figure 2, blue squares). Finally, the glass transition of different DNA samples was observed at temperatures between 205 and 223 K, with an average value of 215±9 K (23–26) (Figure 2, blue triangles). In sum, three terrestrial biopolymers present a glass transition at an average temperature close to 207±16 K (Figure 2, blue horizontal line), very close to the critical temperature of supercooled water at approximately 220±7 K (Figure 1, red horizontal line). From this, we propose that the glass transition temperature of a biopolymer in any liquid can be approximated as the critical temperature of the supercooled liquid.

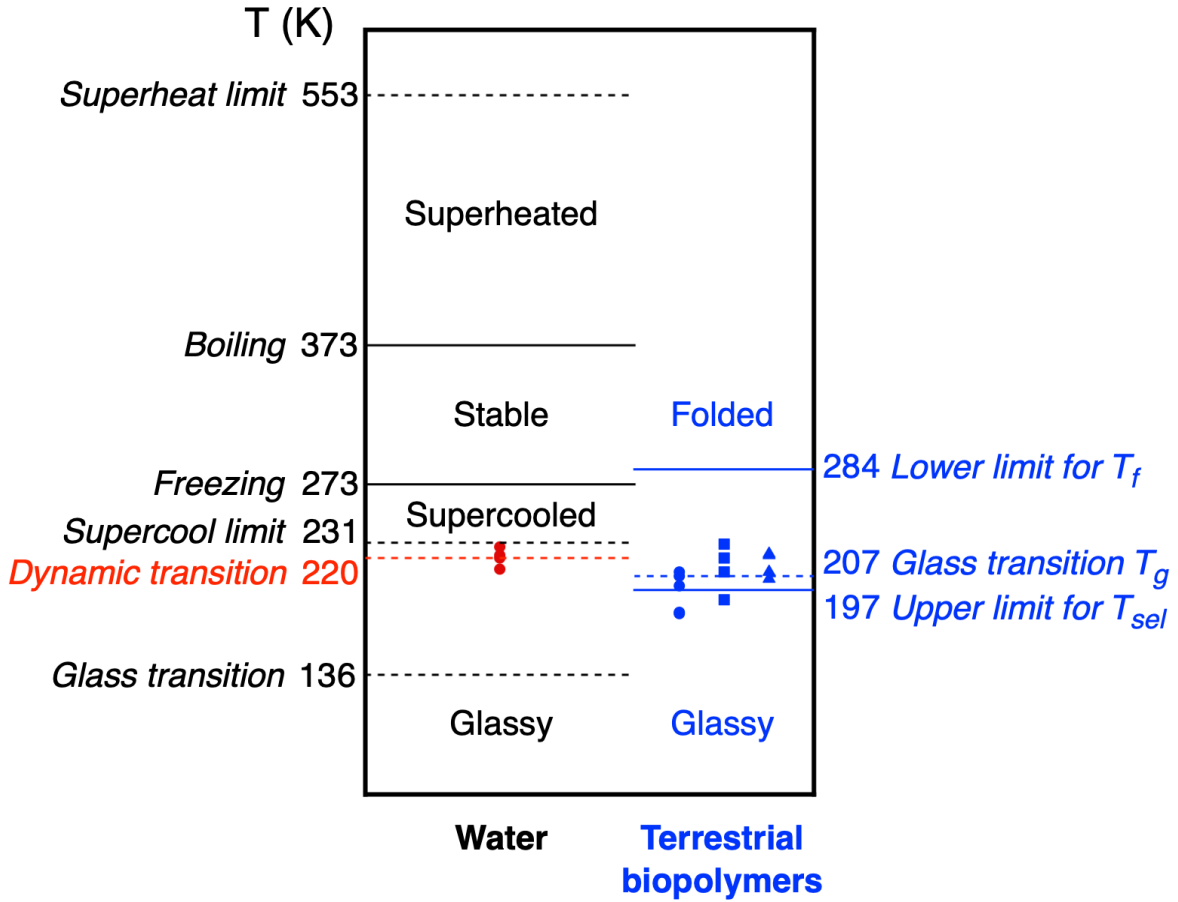

*Figure 2. Temperature domains of water and terrestrial biopolymers at atmospheric pressure. Left: temperature domains of stability and metastability for liquid and glassy water (black) and dynamic transition (red) (11–13). Equilibrium transitions are shown as full lines, kinetically controlled transitions as dashed lines. Right: temperature domains and glass transition for folding of terrestrial biopolymers (blue). Circles indicate proteins (15–19, 27), squares indicate RNA (16, 20–22) and triangles indicate DNA (23–26). Above the glass transition temperature (dashed line) a biopolymer can search for its folded form in an efficient manner. Below the biopolymer glass transition temperature a biopolymer is trapped in a rough energy landscape and will not fold in a biology relevant timescale. The theoretical limits for $T_{sel}$ and $T_f$ calculated from the average glass transition temperature and formula [4] to [6] are shown as continuous lines. Adapted from (11).*

We can put together formulas [4] to [6] relating $T_f$, $T_g$ and $T_{sel}$ for biopolymer folding with our empirical estimation for the glass transition temperature at 207 K. This yields an upper limit for the selection temperature $T_{sel}$ at approximately 197 K (Figure 2, right), which is compatible with previous estimates of $T_{sel}$ close to 108 K (8). It also yields a lower limit for the folding temperature $T_f$ close to 284 K (Figure 1, right), which is compatible with the $T_f$ of most proteins from psychrophilic organisms being above this number (28). We conclude that these theoretical limits for $T_f$, $T_g$ and $T_{sel}$ in water are in reasonable agreement with the available empirical data and can therefore be used as a starting point to analyze biopolymer folding in non aqueous solvents.

**Estimation of the dynamic transition temperature for non-aqueous solvents**

Several solvents other than water have been linked to alternative biochemistries. We evaluate whether they support biopolymer folding within the framework of the presented theory by calculating values of $T_g/T_{sel}$, $T_f/T_g$ and efficiency using equations [7] and [8] and comparing them to the theoretical limits. This requires as a first step the estimation of the dynamic transition temperature for each solvent, which corresponds to $T_g$ for biopolymer folding (Figure 2).

Unfortunately, the dynamic transition temperature remains uncharacterized for most solvents linked to alternative biochemistries. At this point, we recall the existence of linear relationships between the characteristic temperatures of many substances, in relation to the law of corresponding states (29). For example, the ratio between the boiling point and the melting point is roughly constant for nearly 1000 organic substances (29) and for a wide variety of organic and inorganic substances (30). There is also a linear relationship between the glass transition temperature and the melting temperature of metallic glasses (31) and of various polymers (32). Moreover, there are statistically significant linear relationships between the glass transition temperatures, melting points and boiling points of the solvents considered in this work (Table 1 and Figure 3). The correlation between the glass transition temperature and the melting point has a slope of 1.54±0.11 and an intercept of 22±15 K (Pearson correlation coefficient 0.88, p-value 6e-8), while the correlation between the glass transition temperature and the boiling point has a slope of 1.67±0.16 and an intercept of 188±21 K (Pearson correlation coefficient 0.83, p-value 3e-7).

Table 1. Characteristic temperatures of solvents at 1 atm. [a]Calculated from the glass transition temperature and the correlation in Figure 2 except for the upper section of the table. [b]Not available. [c]Fraction of the temperature range between the melting and the boiling point where equations [3] to [6] are fulfilled. [d]Average efficiency for conversion of folding free energy to information in the liquid range between the melting and the boiling point where equations [3] to [6] are fulfilled.

| Substance | Glass transition temperature (K) | Dynamic transition temperature[a] (K) | Melting point (K) | Boiling point (K) | Biopolymer range / liquid range[c] | Average efficiency[d] | Biopolymer Solvent Score | Molecular mass (Da) | References |
|---|---|---|---|---|---|---|---|---|---|
| **Substances with known dynamic transition temperature** | | | | | | | | | |
| Water | 136 | 220 | 273.15 | 373 | 0.88 | 0.57 | 0.852 | 18.0 | (11–13) |
| Ethanol | 97 | 170 | 155.7 | 351.5 | 0.60 | 0.51 | 0.668 | 46.1 | (33, 34) |
| Glycerol | 185 | 310 | 291.2 | 563 | 0.50 | 0.57 | 0.643 | 92.1 | (33, 35) |
| Polybutadiene | 186 | 215 | [b] | [b] | [b] | [b] | [b] | [b] | (36) |
| m-tricresyl phosphate | 210 | 260 | 233 | 528 | 0.58 | 0.52 | 0.659 | 368.4 | (35) |
| Polycholoro trifluoro ethylene | 213 | 318 | [b] | [b] | [b] | [b] | [b] | [b] | (36) |
| Phenyl salicylate | 218 | 256 | 315 | 446 | 0.72 | 0.59 | 0.779 | 214.2 | (35) |
| Ortoterphenyl | 243 | 291 | 486 | 662 | 1.00 | 0.41 | 0.773 | 230.3 | (36) |
| 0.4 Ca(NO$_3$)$_2$ 0.6 KNO$_3$ | 333 | 367 | [b] | [b] | [b] | [b] | [b] | 126.3 | (36) |
| Tri-alpha-naphthyl benzene | 342 | 424 | [b] | [b] | [b] | [b] | [b] | 456.6 | (36) |
| Polystyrene | 373 | 445 | [b] | [b] | [b] | [b] | [b] | [b] | (36, 37) |
| **Hydrocarbons** | | | | | | | | | |
| Ethane | 46 | 124 | 90.4 | 184.6 | 0.14 | 0.65 | 0.373 | 46.1 | (38) |
| 2-methylbutane | 69 | 146 | 113.3 | 306.0 | 0.54 | 0.51 | 0.632 | 72.2 | (33) |
| 2-methyl-2-butene | 73 | 150 | 149 | 311.4 | 0.65 | 0.51 | 0.694 | 70.1 | (33) |
| 2-methylpentane | 79 | 155 | 119.5 | 333.4 | 0.56 | 0.50 | 0.637 | 86.2 | (33) |
| 3-methylhexane | 88 | 164 | 153.6 | 365 | 0.66 | 0.49 | 0.679 | 100.2 | (33) |
| Methylcyclohexane | 86 | 162 | 146.6 | 373.3 | 0.66 | 0.47 | 0.672 | 98.2 | (33) |
| 4-methylcyclohexene | 94 | 169 | 157.6 | 375.7 | 0.66 | 0.49 | 0.677 | 96.2 | (33) |
| Ethylcyclohexane | 98 | 173 | 141.3 | 404.9 | 0.63 | 0.47 | 0.652 | 112.2 | (33) |
| Isopropylcyclohexane | 108 | 183 | 183.3 | 427.6 | 0.72 | 0.47 | 0.695 | 126.2 | (33) |
| n-butylcyclohexane | 119 | 193 | 198.4 | 454.0 | 0.73 | 0.46 | 0.701 | 140.3 | (33) |
| **Small substances from C, H, O, N and S** | | | | | | | | | |
| Methanol | 103 | 178 | 175.2 | 337.6 | 0.57 | 0.55 | 0.673 | 46.1 | (33) |
| 1-propanol | 102 | 177 | 146.6 | 370.1 | 0.57 | 0.51 | 0.645 | 60.1 | (33) |
| 1-butanol | 115 | 189 | 183.6 | 390.3 | 0.63 | 0.51 | 0.685 | 74.1 | (33) |
| Ethylene glycol | 155 | 227 | 255.6 | 470.2 | 0.74 | 0.51 | 0.740 | 62.1 | (33) |
| Cyclohexanol | 161 | 232 | 297 | 434.5 | 0.84 | 0.56 | 0.821 | 100.2 | (33) |
| Acetaldehyde | 82 | 158 | 149.5 | 294 | 0.53 | 0.56 | 0.657 | 44.1 | (33) |
| Ether | 91 | 167 | 156.7 | 307.6 | 0.52 | 0.56 | 0.647 | 74.1 | (33) |
| Acetone | 96 | 171 | 178 | 329.5 | 0.63 | 0.54 | 0.700 | 58.1 | (33) |
| Formic acid | 148 | 220 | 281.5 | 373.9 | 0.77 | 0.60 | 0.815 | 46.0 | (38) |
| Hydrazine | 128 | 201 | 275 | 387 | 0.99 | 0.54 | 0.883 | 32.0 | (38) |
| Methylamine | 93 | 168 | 180 | 267 | 0.41 | 0.63 | 0.612 | 31.1 | (38) |
| Acetonitrile | 93 | 168 | 232 | 335 | 1 | 0.53 | 0.873 | 41.1 | (33) |
| Dimethylformamide | 129 | 202 | 212 | 426 | 0.69 | 0.51 | 0.711 | 73.1 | (33) |
| Carbon disulfide | 92 | 167 | 161.5 | 319.4 | 0.57 | 0.55 | 0.669 | 76.1 | (39) |
| Dimethyl sulfoxide | 150 | 222 | 291.6 | 462.1 | 0.92 | 0.51 | 0.827 | 78.1 | (33) |
| Sulfuric acid | 160 | 231 | 283 | 573 | 0.88 | 0.44 | 0.751 | 98.1 | (40) |
| Dimethyl sulfone | 190 | 260 | 383 | 511 | 1.00 | 0.51 | 0.858 | 94.1 | (33) |
| **Halogenated substances** | | | | | | | | | |
| Methylene chloride | 101 | 176 | 176.3 | 313.1 | 0.52 | 0.58 | 0.657 | 84.9 | (33) |
| Chloroform | 110 | 184 | 209.5 | 334 | 0.65 | 0.57 | 0.732 | 119.4 | (33) |
| Carbon tetrachloride | 130 | 203 | 250.2 | 349.8 | 0.71 | 0.59 | 0.778 | 153.8 | (33) |
| Stannic chloride | 130 | 203 | 240 | 387.1 | 0.73 | 0.55 | 0.763 | 260.5 | (33) |
| 2-bromobutane | 97 | 172 | 161.1 | 364.3 | 0.63 | 0.50 | 0.675 | 137.0 | (33) |
| Chlorobenzene | 127 | 200 | 228 | 405 | 0.73 | 0.52 | 0.745 | 112.6 | (33) |
| Bromobenzene | 137 | 210 | 242.4 | 428 | 0.75 | 0.52 | 0.749 | 157.0 | (33) |
| **Aromatic substances** | | | | | | | | | |
| Benzene | 131 | 204 | 278.5 | 353.1 | 0.96 | 0.59 | 0.903 | 78.1 | (33) |
| Toluene | 116 | 190 | 178 | 383 | 0.60 | 0.53 | 0.672 | 92.1 | (33) |

| Compound | | | | | | | | | |
|---|---|---|---|---|---|---|---|---|---|
| Ethylbenzene | 112 | 186 | 179.1 | 409.1 | 0.67 | 0.49 | 0.659 | 106.2 | (33) |
| n-propylbenzene | 124 | 198 | 173.6 | 432.3 | 0.62 | 0.49 | 0.665 | 120.2 | (33) |
| n-butylbenzene | 127 | 200 | 191.8 | 456.3 | 0.69 | 0.48 | 0.686 | 134.2 | (33) |
| tert-butylbenzene | 140 | 213 | 214.9 | 442.1 | 0.66 | 0.51 | 0.697 | 134.2 | (33) |
| n-pentylbenzene | 132 | 205 | 198.1 | 478.5 | 0.70 | 0.47 | 0.687 | 148.2 | (33) |
| Phenol | 202 | 271 | 314 | 455 | 0.58 | 0.60 | 0.711 | 94.1 | (33) |
| Anisole | 122 | 196 | 235.7 | 428 | 0.82 | 0.49 | 0.765 | 108.1 | (33) |
| Benzyl alcohol | 171 | 242 | 257.7 | 478.2 | 0.66 | 0.53 | 0.712 | 108.1 | (33) |
| Benzaldehyde | 148 | 220 | 247.0 | 452.5 | 0.73 | 0.52 | 0.736 | 106.1 | (33) |
| Pyridine | 122 | 196 | 231.3 | 388.7 | 0.75 | 0.53 | 0.760 | 79.1 | (33) |
| Nitrobenzene | 161 | 232 | 244.6 | 374.3 | 0.43 | 0.62 | 0.618 | 123.1 | (33) |
| Aniline | 190 | 260 | 266.8 | 457.4 | 0.52 | 0.58 | 0.662 | 93.1 | (33) |

The upper section of Table 1 compiles the glass transition and dynamic transition temperatures of water and ten non-aqueous liquids at 1 atm. Figure 3C shows that there is a statistically significant correlation between the glass transition and dynamic transition temperatures of these substances (Pearson correlation coefficient 0.94, p-value 0.02), with a slope of 0.94±0.12 and an intercept of 81±28 K. We conclude that we can use this linear relationship to estimate the dynamic transition temperature from the glass transition temperature for other substances with reasonable confidence for a glass transition temperature in the 97-373 K range.

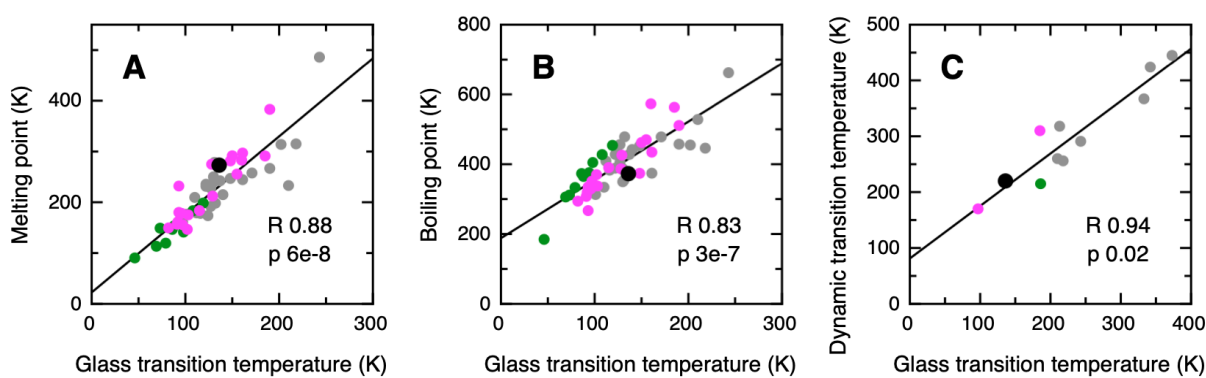

*Figure 3. Characteristic temperatures of solvents. Correlations between the glass transition temperature and (A) the melting point (B) the boiling point (C) the dynamic transition temperature for substances in table 1 at 1 atm. Water is shown as a black circle, hydrocarbons as green circles lines, small substances from C, H, O, N and S as magenta circles and the remaining solvents in Table 1 as grey circles.*

**Characteristic temperatures of solvents linked to alternative biochemistries**

We include in this study a total of 56 solvents from four broad categories (Table 1). The first category is water and other solvents for which the dynamic transition temperature has been characterized. The second category are hydrocarbons, which have been linked to alternative biochemistries (41, 42). The third category are low molecular weight (up to approximately 100 Da) substances made out of elements abundant in the universe such as H, C, N, O and S (43), which may potentially partake in alternative biochemistries. The fourth category are other substances for which we know the melting point, boiling point and the glass transition temperature, such as halogenated substances and aromatic substances. These are included in order to explore a wide range of solvent physical properties.

The present approach requires knowledge of the melting point, boiling point and dynamic transition temperature of each solvent. The values for the melting point and boiling point are taken from the literature. Since the dynamic transition temperature of most relevant solvents remains uncharacterized, we calculate the dynamic transition temperature for the remaining solvents from literature values for the glass transition temperature and the linear relationships shown in Figure 3. We restrict ourselves to substances with a known glass transition temperature in the range of 40-400 K to avoid large extrapolations in the calculations. These requirements leave out of consideration liquids without a documented glass transition temperature in the 40-400 K range, such as ammonia, formamide, hydrogen sulfide, dinitrogen and dihydrogen. Last, we exclude supercritical fluids, where applicability of the theory is unclear since their properties lie somewhere between a liquid and a gas.

**Biopolymer folding in alternative solvents**

We use equations [3], [4] and [8] to calculate the $T_f/T_g$ ratio, the efficiency for conversion of free energy into information and the $T_g/T_{sel}$ ratio as a function of $T_f$, for 54 solvents, for the temperature range in which each substance is a liquid at 1 atm. The results are shown in Figure 4. Water is shown as a thick black line, hydrocarbons as green lines, small substances from C, H, O, N and S as magenta lines and the remaining solvents in Table 1 as thin black lines. The horizontal lines highlight the theoretical limits for each parameter, with those from energy landscape theory shown as continuous lines and those from molecular information theory as dashed lines.

The ratio $T_f/T_g$ increases linearly with $T_f$, the efficiency decreases monotonically with $T_f$ and the ratio $T_g/T_{sel}$ increases monotonically with $T_f$ in the biologically relevant range (Figure 4). The values of $T_f$ for which the $T_f/T_g$ ratio and the efficiency cross their theoretical limits of approximately 1.373 and 0.69 are the same in all panels due to the linkage between equations [2] and [3] and coincide with the highest value of $T_f$ for which the $T_g/T_s$ ratio crosses its theoretical limit of 1.048. Above this value of $T_f$, foldable biopolymer sequences can be coded and evolve in a relevant timescale, and evolved biopolymer sequences can fold in a relevant timescale. All solvents considered have $T_f/T_g$ and $T_g/T_{sel}$ ratios above the theoretical minima and an efficiency below the theoretical maximum for at least a fraction of the temperature range in which they are liquid (Figure 4). Thus, all substances considered here are in principle compatible with the existence of some biopolymer at 1 atm.

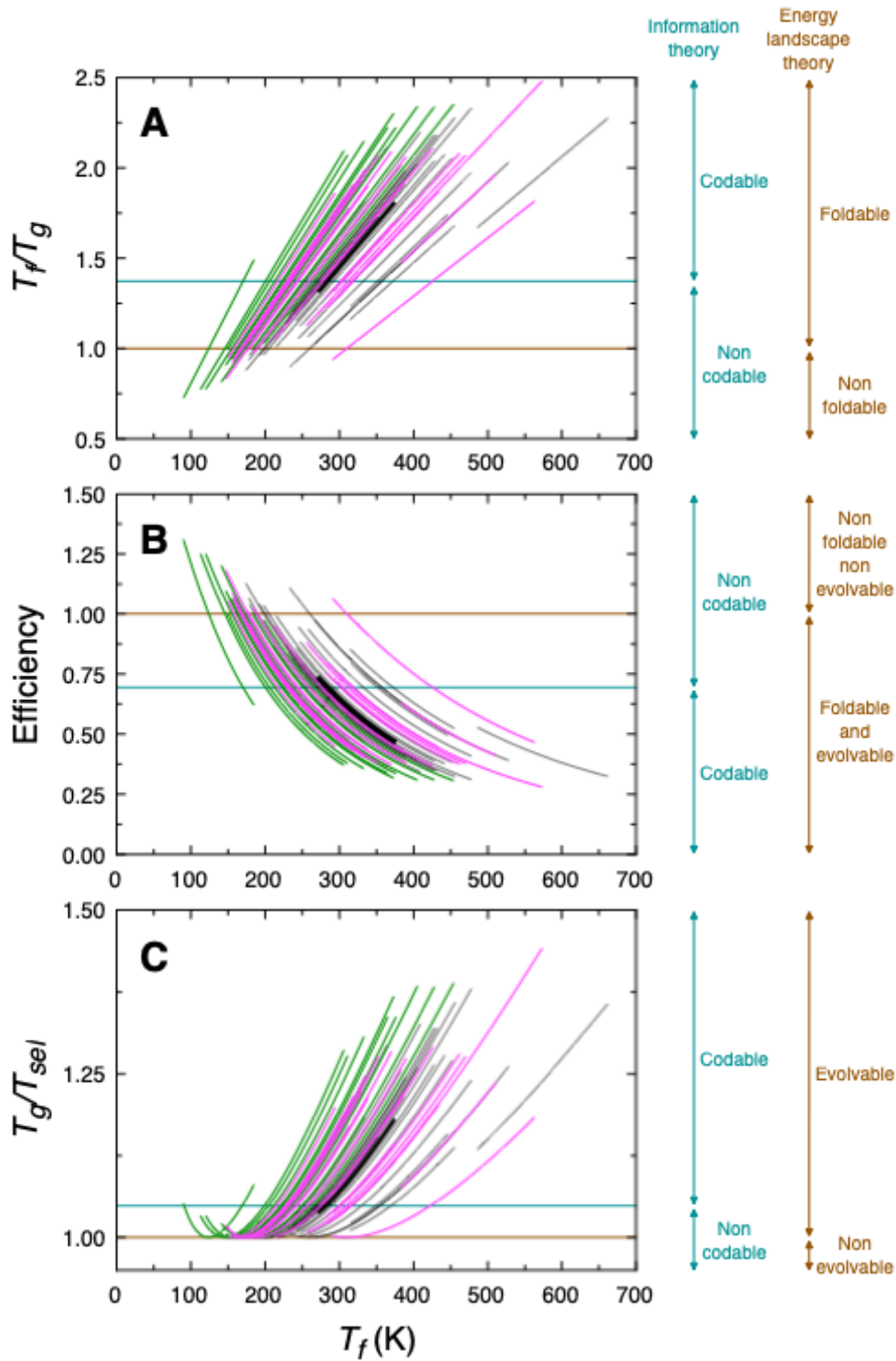

*Figure 4. Temperature dependence of parameters relevant to biopolymer foldability, codability and evolvability, for 54 solvents at 1 atm.* Water is shown as a thick black line, hydrocarbons as green lines, small substances from C, H, O, N and S as magenta lines and the remaining solvents in Table 1 as grey lines. The horizontal lines highlight the theoretical limits for each parameter, with those from energy landscape theory shown as blue lines and those from molecular information theory in brown lines. (A) $T_f/T_g$ ratio. (B) Efficiency for conversion of folding free energy into sequence information. (C) $T_g/T_{sel}$ ratio.

**Is water particularly well-suited for biopolymer folding?**

We present a quantitative comparison between the substances in Table 1 with regard to their support for biopolymer folding and evolution, with particular attention to water. Figure 4 shows that all solvents considered have values of $T_f$, $T_g$, $T_{sel}$ and efficiency compatible with biopolymer folding and evolution for at least a fraction of the temperature range in which they are liquid (Figure 4). We focus here on the efficiency for conversion of folding free energy into sequence information, which takes values between 0.25 and 0.69, and on the fraction of the liquid range that is compatible with folding, which ranges from 0.14 to 1. In principle, one may expect that higher values for both the average efficiency and the fraction of the liquid range compatible with folding and evolution are preferable. We can normalize the average efficiency by its theoretical maximum $ln(2)$ so that both magnitudes range between 0 and 1 and define a Biopolymer Solvent score (BSscore) for each solvent that is the geometric mean of these:

$$BSscore = \sqrt{\frac{Average\ Efficiency}{Maximal\ efficiency} \cdot \frac{Biopolymer\ Range}{Liquid\ Range}} \quad [9]$$

Figure 5A shows the relationship between the two variables for 54 substances in Table 1 at 1 atm. Water is shown as a black circle, hydrocarbons as green circles, small substances from C, H, O, N and S as magenta circles and the remaining solvents in Table 1 as grey circles. The grey lines indicate constant values of the *BSscore*. Since Figure 2 shows considerable uncertainty in the dynamic transition temperature of water, with values ranging from 180 to 230K, we recalculated the *BSscore* for water using values within this range. This yields the black line in Figure 5, with *BSscore* values from 0.71 to 0.89 (average 0.82±0.06), confirming that water belongs to the group of solvents with both a high average efficiency for conversion of folding free energy into sequence information and a large fraction of the liquid range compatible with folding. As suspected, water appears to be an exceptionally good solvent to support biopolymer folding and evolution. The remaining solvents can be ranked by the *BSscore* and the distribution is shown in Figure 5B.

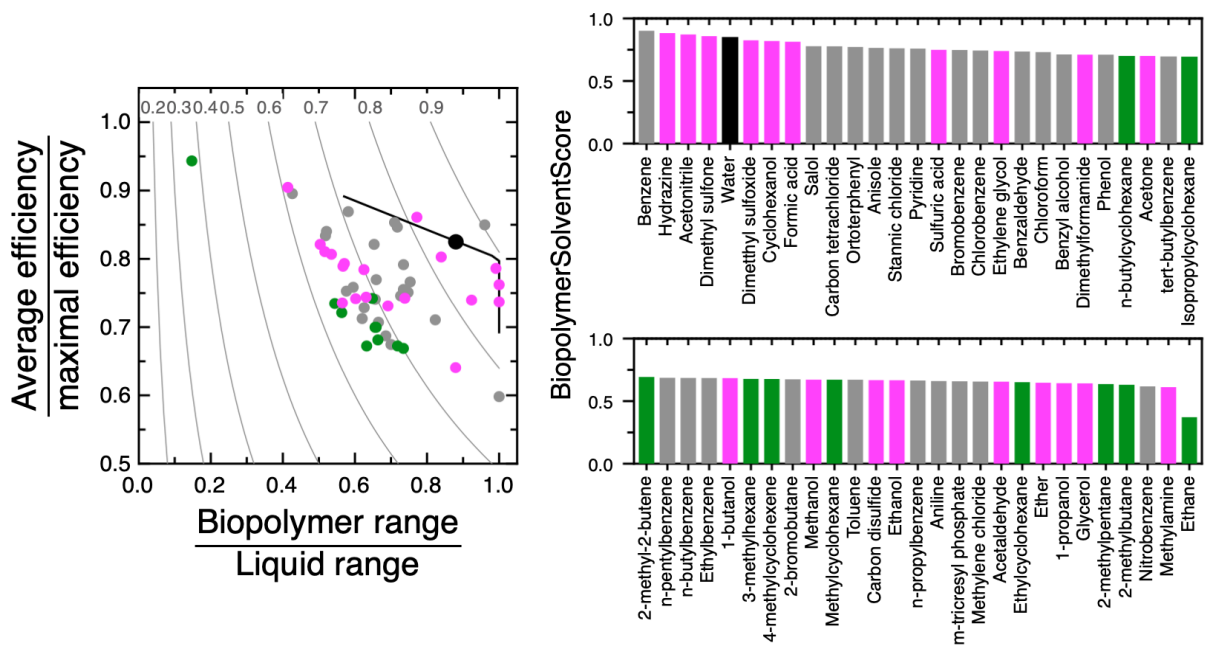

*Figure 5. Biopolymer Solvent scoring.* (A) *Relationship between the fraction of the biopolymer compatible range and the liquid range of solvents and the average fractional efficiency for conversion of folding free energy to sequence information, for 54 solvents at 1 atm. Large black circle: water. Green circles: hydrocarbons. Magenta circles: small substances from C, H, O, N and S. Grey circles: other solvents in Table 1. Black line: range for water calculated from the uncertainty in the dynamic transition temperature presented in Figure 3 (values range from 180 to 230K). The grey lines indicate constant values of the BSscore (Equation [9]).* (B) *Ranking of the solvents according to the BSscore.*

**Role of pressure**

The presented results suggest that all 54 solvents analyzed may support biopolymer folding and evolution at 1 atm, at least in some range of temperatures. However, solvents may be subject to very different pressures in extraterrestrial environments. For example, the atmospheric pressures in the solar system may range from tens of microbars to nearly 100 bar (44). Liquid water can exist in a large fraction of this pressure range, from approximately $6 \times 10^{-3}$ to 220 bar. It seems relevant to ask how the results presented in Figures 4 and 5 change at different pressures. The temperature for the dynamic transition of water and for its melting point vary little with pressure (13, 45). On the other hand, the boiling point of water increases sharply with increasing pressure (45). We expect other solvents to behave similarly due to the small changes in molar volume associated with the glass transition and boiling, relative to the change in molar volume upon boiling (46).

We may now reconsider our calculations with $T_g$ and the melting point as constants and a boiling point that increases with pressure. The main effect would then be an increase of the liquid range from the right extreme of all lines in Figure 4. The effect of this would be two-fold: the average efficiency for conversion of folding free energy into sequence information would decrease and the fraction of the liquid range where folding is possible would increase. The outcome in terms of *BSscore* would depend on the interplay between these two factors. In the case of water,

Figure 6 shows the pressure dependence of the *BSscore* (black), the normalized average efficiency (blue) and the fraction of the liquid range where folding is possible (orange). In this case, the efficiency is maximal at lower pressures, the fraction of the liquid range where folding is possible is maximal at higher pressures and the *BSscore* varies within a relatively narrow range (0.45 to 0.86), is above 0.7 for pressures above 0.04 atm, and is close to the maximum value (0.85) at 1 atm. Overall, the influence of pressure on the ability of water to support biopolymer folding is modest. Since the overall shape of the phase diagram is expected to be the same for other substances, we propose this conclusion holds for the other solvents considered here.

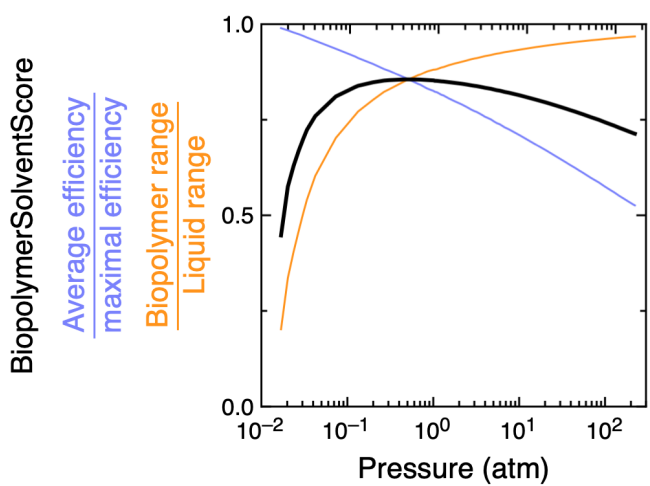

*Figure 6. Pressure dependence for biopolymer folding and evolution in water. Pressure dependence of the BSscore (black), the normalized average efficiency (blue) and the fraction of the liquid range where folding and evolution are possible for water (orange).*

**Discussion**

We are still uncertain if Life happens in other parts of the universe besides our beloved planet Earth. Much research and lots of speculations are being put forward yet few consensus on how Life may look like, and thus how to detect it have been reached (47). At the molecular level, is life the same everywhere? Are there alternative biochemistries? To gain insight into these questions we propose to abstract out the many known details of terrestrial biochemistry and ask: What are physicochemical characteristics of the solvents that may sustain life that constrain biopolymer folding and evolution?

Energy landscape and molecular information theories allow one to put the fundamentals of biopolymer folding, coding and evolution under scrutiny given the assumption that these must happen in liquid solvents and some sequence space. We have shown that the physical characteristics of solvents constrain the ranges where folding and evolution are possible. Folding is constrained by the characteristic temperatures of the solvents, and coding is constrained by the efficiency of the energy-to-information conversion. We believe that both must play a significant role in determining the emergence of biopolymers, regardless of their detailed chemical composition.

The ability of most nonaqueous solvents to support biopolymer folding in general is in line with previous experimental results with specific terrestrial biopolymers. For example, several DNA duplexes maintain their structure in glycerol and ethylene glycol (48), and the proteins subtilisin and lysozyme remain folded in glycerol, dioxane and acetonitrile (49–52). Lysozyme can undergo reversible folding in glycerol (53). Moreover, many enzymes retain their catalytic activities in nonaqueous organic solvents (54–56). It seems likely that most mixtures of the solvents considered here also support biopolymer folding. This opens the door to considering the existence of life in mixed solvents.

We can speculate on whether terrestrial biopolymers (proteins, RNA, DNA) can evolve, fold and function in any non aqueous solvent. Our results suggest that the solvent requirements of evolvability and foldability can be fulfilled in most cases. However, it should also be kept in mind that biopolymer folding and function also require that the folded state is more stable than the unfolded state and soluble in the new solvent (57). Terrestrial proteins are likely to be less stable in polar solvents such as ethanol, and markedly insoluble in non-polar solvents such as cyclohexane. This is clearly due to the physicochemical properties of the genetically coded terrestrial amino acids in relation to the properties of the solvent (57). Thus, it seems unlikely that the current set of terrestrial amino acids can support life in solvents other than water (57).

Life in nonaqueous solvents would likely be supported by biopolymers that are chemically different from the terrestrial ones. There are three main points relevant to the suitability of a protein-like biopolymer to evolve, fold and sustain life in a given solvent. We can first consider that biopolymer sequences must evolve to the point in which they contain enough information to specify a given fold (6). This minimal amount of information is determined by the configurational entropy change upon folding of the biopolymer in a given solvent (6). The evolutionary choice of a given biopolymer-solvent pair poses an informational requirement for folding. Second, folding of a biopolymer in a given solvent is associated with a favorable change in the free energy of the system. At the atomic scale, folding involves breaking a myriad of molecular interactions between solvent molecules and between solvent and biopolymer and forming precise intramolecular interactions in the biopolymer. The free energy balance associated with this process is also dictated by the properties of the solvent-biopolymer pair (57). Third, the properties of the solvent determine the efficiency for conversion of free energy into sequence information. Interestingly, the efficiency expected for different solvents is predicted to be at least half of the maximal one in most solvents examined (Figure 5). The maximal amount of information that is gained by a biopolymer by folding in a solvent is dictated by the properties of both solvent and biopolymer. If a biopolymer is to evolve and fold in a certain solvent, the amount of information gained by conversion of folding free energy should match the amount of information required to specify its folded conformation (6). Both of these quantities are determined by physicochemical properties of the solvent-biopolymer pair. We speculate that biopolymers associated with alternative solvents must be chemically different from those found in water-based terrestrial life. Alternative chemical structures previously suggested for biopolymers include polyelectrolytes in general (58), protein-like polyesters (59, 60), peptides made from beta- and gamma-amino acids (61) and polymers originating in the opening of common cyclic prebiotic chemicals (62).

The theory presented here is universal for biopolymers that fold and function in regular fluids, those for which we can estimate the dynamic transition temperature from the glass transition temperature. Future work may consider similar questions in solvents without a known dynamic or glass transition in the 40-400 K range, such as ammonia (41), in supercritical fluids such as carbon dioxide (63), for biopolymer folding *in vacuo* (64), for biopolymers that fold in a downhill manner (65), for biopolymers that function through liquid-liquid phase separation (66), and for intrinsically disordered biopolymers (67).

All solvents considered here support biopolymer folding under some conditions, with efficiencies for conversion of free energy into information in the same order of magnitude (Figure 4B). This result can be traced back to the correlations between the characteristic temperatures of these substances (Figure 3), which can be explained by the law of corresponding states. This law recasts the properties of fluids in terms of their critical points. In this scenario, the properties that vary among fluids determine their critical temperature and pressure. The degree to which all fluids deviates from ideal gas behavior is the same at the same reduced temperature and pressure, which in turn leads to the correlations between the characteristic temperatures (29). The law of corresponding states assumes that intermolecular interactions in the fluid are a function only of the ratio between the intermolecular distance $R$ and a characteristic distance $R_0$ (68). This assumption is accurate for large values of $R$ where the intermolecular potential energy is proportional to $R^{-6}$ (68). To a point, our results can then be traced back to the existence of Van der Waals interactions, which is undoubtedly a universal phenomenon. Substances which are not spherically symmetrical and for which other intermolecular forces are relevant deviate to varying degrees from the law of corresponding states, leading to deviations from the correlations between characteristic temperatures and to the diversity of behaviors shown in Figure 4. Thus, the differences in efficiency and in the range of liquid temperatures where a solvent supports folding and evolution are fundamentally related to the rotational asymmetry and polarity of its constituent molecules. We have shown here how to derive the basic characteristics expected for biopolymers from these fundamental principles.

**Acknowledgements**

This work was supported by the Consejo de Investigaciones Científicas y Técnicas (CONICET) (IES and DUF are CONICET researchers and EAG is a postdoctoral fellow); CONICET Grant PIP2022-2024 - 11220210100704CO. Additional support from NAI and Grant Number 80NSSC18M0093 Proposal ENIGMA: EVOLUTION OF NANOMACHINES IN GEOSPHERES AND MICROBIAL ANCESTORS (NASA ASTROBIOLOGY INSTITUTE CYCLE 8).